\documentclass[12pt]{article}
\pdfoutput=1
\usepackage{graphicx}
\usepackage{subfigure}

\def\pbnr{}
\def\speaker{Andrea Contu}
\def\onbehalfof{The LHCb collaboration}
\def\title{Status and prospects for the LHCb upgrade}
\def\affiliation{INFN Sezione di Cagliari, Italy\\
and CERN, Switzerland}
\def\support{The workshop was supported by the University of Manchester, IPPP, STFC, and IOP}

\newcommand\pubnumber{\pbnr}
\newcommand\pubdate{\today}
\def\Title#1{\begin{center} {\Large #1 } \end{center}}
\def\Author#1{\begin{center}{ \sc #1} \end{center}}

\newcommand{\OnBehalf}[1]{\sbox0{#1}\ifdim\wd0=0pt
        {}
	\else
	{\\on behalf of #1}
	\fi}
\newcommand{\SupportedBy}[1]{\sbox0{#1}\ifdim\wd0=0pt
        {}
	\else
	{\footnote{#1}}
	\fi}
\def\Address#1{\begin{center}{ \it #1} \end{center}}

\newcommand\pubblock{\includegraphics[width=5cm]{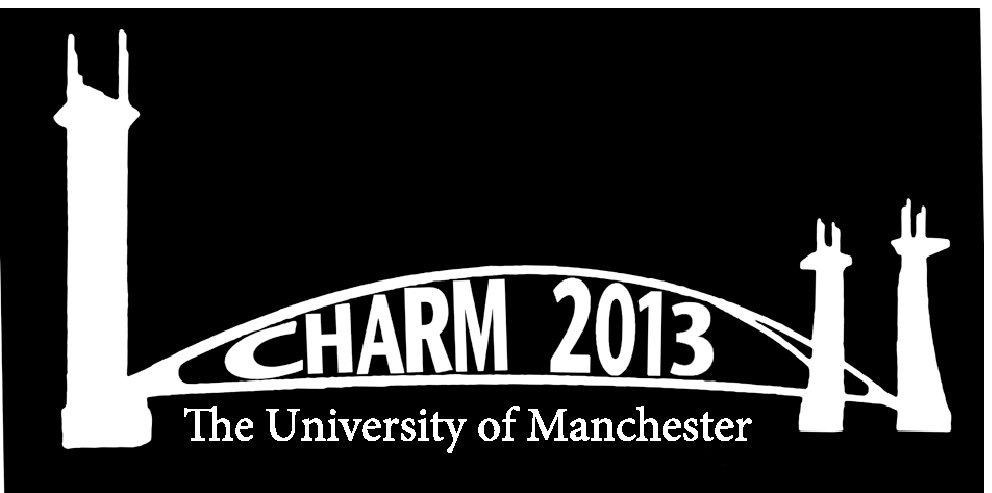}\hfill{\begin{tabular}{l} \pubnumber\\
         \pubdate  \end{tabular}}}
\newenvironment{Abstract}{\begin{quotation}  }{\end{quotation}}
\newenvironment{Presented}{\begin{quotation} \begin{center} 
             PRESENTED AT\end{center}\bigskip 
      \begin{center}\begin{large}}{\end{large}\end{center} \end{quotation}}

\def\venue{The 6$^{\mathrm{th}}$ International Workshop on Charm Physics\\
(CHARM 2013)\\
Manchester, UK,  31 August -- 4 September, 2013}

\def\beq{\begin{equation}}
\def\eeq#1{\label{#1}\end{equation}}
\def\eeqn{\end{equation}}

\def\beqa{\begin{eqnarray}}
\def\eeqa#1{\label{#1}\end{eqnarray}}
\def\eeqan{\end{eqnarray}}

\let\bar=\overbar

\def\Dslash{\not{\hbox{\kern-4pt $D$}}}
\def\dslash{\not{\hbox{\kern-2pt $\del$}}}

\def\msb{{\bar{\ssstyle M \kern -1pt S}}}

\begin{document}
\begin{titlepage}
\pubblock

\vfill
\Title{\title}
\vfill
\Author{\speaker\SupportedBy{\support}\OnBehalf{\onbehalfof}}
\Address{\affiliation}
\vfill
\begin{Abstract}
High-precision measurements performed by the LHCb collaboration have opened a new era in charm physics. Several crucial measurements, particularly in spectroscopy, rare decays and $CP$ violation, can benefit from the increased statistical power of an upgraded LHCb detector. The upgrade of LHCb detector, its software infrastructure, and the impact on charm physics are discussed in detail.
\end{Abstract}
\vfill
\begin{Presented}
\venue
\end{Presented}
\vfill
\end{titlepage}
\def\thefootnote{\fnsymbol{footnote}}
\setcounter{footnote}{0}
%

\section{Introduction}
The LHC has performed excellently during its first years of operation allowing the four main experiments to collect large data samples at unprecedented centre-of-mass energies. The LHCb detector outperformed its design specification and played a crucial role in the advancement of charm physics. The LHCb  measurements range from the charm cross-section at $\sqrt{s}=7\,\mathrm{TeV}$ \cite{Aaij:2013mga}, to direct and indirect $CP$ violation, neutral charm meson mixing, spectroscopy, and rare decays. These measurements exploit the large charm cross-section at the LHC and the outstanding performance of the trigger and reconstruction system of LHCb, which allowed unprecedented charm yields to be available for precision analyses. Charm physics plays a crucial role in the LHCb upgrade programme as well, in which the sensitivity for several key observables is expected to reach or exceed the theoretical precision. 
In this paper, the LHCb upgrade, both from the hardware and software point of view, is outlined. Prospects for charm physics in the LHCb upgrade era are discussed and extrapolations of the expected sensitivities for several observables are listed.

The scientific value of these advances has been recognised by the CERN research board, which approved the upgrade of LHCb to be part of the long-term exploitation of the LHC.

\section{The LHC upgrade schedule}

The first running phase of the LHC, with $pp$ centre of mass energy of 7 and 8 TeV, ended at the beginning of 2013. Currently, the LHC machine and the four experiments are in a 18-months shutdown (LS1) for maintenance and consolidation. Data taking will be resumed at the beginning of 2015 with a $pp$ center of mass energy of $13\textendash14\,\mathrm{TeV}$.  The spacing between consecutive proton bunches circulating in the accelerator is foreseen to go from $50\,\mathrm{ns}$ to the nominal $25\,\mathrm{ns}$, effectively doubling the $pp$ collision rate. From the beginning of 2018 a second long shutdown (LS2) is expected to last about a year, followed by three years of running up to 2022, after which a luminosity upgrade of the LHC is foreseen. It is noted that this schedule is likely to evolve with time.

\section{The current LHCb detector and its upgrade}
The LHCb detector~\cite{Alves:2008zz} is a single-arm forward
spectrometer covering the pseudorapidity range $2<\eta <5$,
designed for the study of particles containing $b$ or $c$
quarks. The detector includes a high-precision tracking system
consisting of a silicon-strip vertex detector surrounding the $pp$
interaction region, a large-area silicon-strip detector located
upstream of a dipole magnet with a bending power of about
$4{\rm\,Tm}$, and three stations of silicon-strip detectors and straw
drift tubes placed downstream.
The combined tracking system provides a momentum measurement with
relative uncertainty that varies from 0.4\% at 5 $\mathrm{GeV}/c$ to 0.6\% at 100 $\mathrm{GeV}/c$,
and impact parameter resolution of 20 $\mathrm{\mu m}$ for
tracks with large transverse momentum. Different types of charged hadrons are distinguished by information
from two ring-imaging Cherenkov detectors~\cite{LHCb-DP-2012-003}. Photon, electron and
hadron candidates are identified by a calorimeter system consisting of
scintillating-pad and preshower detectors, an electromagnetic
calorimeter and a hadronic calorimeter. Muons are identified by a
system composed of alternating layers of iron and multiwire
proportional chambers~\cite{LHCb-DP-2012-002}.

The upgraded LHCb detector is expected to be installed in 2018, during LS2, and is currently being designed to perform as well as or better than the current one at a higher instantaneous luminosity.
The physics goal for the upgrade is to reach a sensitivity at the level of the theoretical prediction (or better) in several key observables. Therefore, in order to keep the same level of performance in harsher conditions, improvements in the trigger, reconstruction strategy, and detector technology are mandatory.
The total integrated luminosity collected at the end of the LHCb upgrade data taking is expected to reach $70\,\mathrm{fb^{-1}}$.

\subsection{Trigger strategy}
\begin{figure}[!ht]
\centering
 \subfigure[Current trigger system]{\includegraphics[width=5cm]{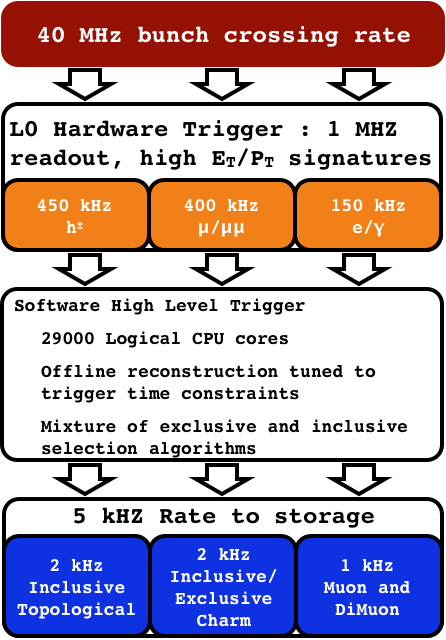}\label{fig:trigger1a}}
{\hspace{2cm}}
\subfigure[Trigger system in the upgrade]{\includegraphics[width=5cm]{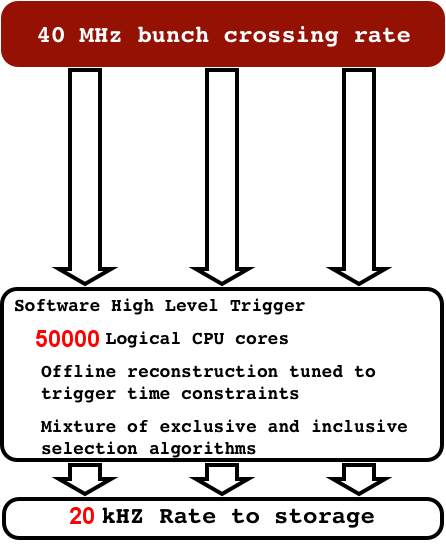}\label{fig:trigger1b}}
\caption{Overview of current and planned LHCb trigger system.}
\label{fig:trigger1}
\end{figure}

\begin{figure}[!ht]
\centering
 \includegraphics[width=8cm]{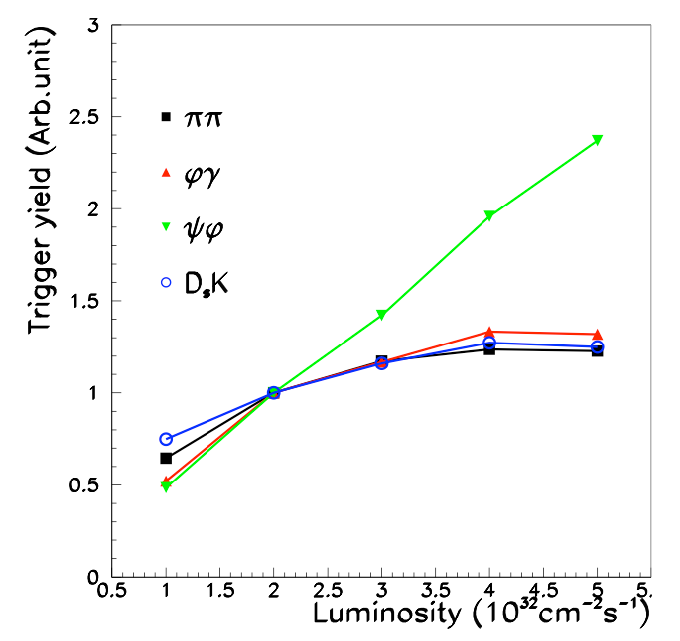}
\caption{Trigger yield for several $B$ decays as a function of the instantaneous luminosity in the current trigger scheme. $B^0\to\pi^+\pi^-$ is represented by black squares, $B^0\to \phi \gamma$ by red triangles, $B^0_s\to J/\psi \phi$ by green upside-down triangles and $B_s^0\to D_s^+K^-$ by blue circles.}
\label{fig:limitations}
\end{figure}

The current trigger scheme is based on a multi-stage approach with a first level, hardware-based, trigger and two software levels that have access to the full event information (see \figurename{~\ref{fig:trigger1a}}).
The output rate of the first hardware-based trigger level, which uses information on transverse momentum, $p_T$, and transverse energy $E_T$, is limited by a maximum bandwidth of 1.1 MHz.
At higher luminosity, this constraint would require using tighter $p_T$ and $E_T$ cuts in hadronic triggers in order for the computing infrastructure to cope with the increased event rate and size. This will also cause the trigger efficiency for hadronic channels to deteriorate, as shown in \figurename{~\ref{fig:limitations}}. On the other hand, events that are selected by muonic triggers will be mostly unaffected since the muon system is already capable of sustaining a higher instantaneous luminosity to some extent.
The effect is even more pronounced for charm hadrons, which are produced at a lower $p_T$ than $b$ hadrons.
The inefficiency in for the hadronic triggers will also affect the charm yield achievable. 

A new trigger strategy for the upgrade is being studied in which the first level hardware trigger is completely removed and the events are sent directly to a software trigger running on a larger and more powerful CPU farm, as shown in \figurename{~\ref{fig:trigger1b}}. This new scheme is not affected by the ``bandwidth bottleneck'' after the first trigger level so that the event rate that can be processed and stored on disk depends only on the capabilities of the CPU farm. The final event output rate is expected to  be a factor of four larger than the current one.

\subsection{Tracking system and RICH upgrade}
One of the obvious effects of the increased instantaneous luminosity is a higher occupancy and radiation dose for all the subdetectors. Layout and technology improvements are needed to cope with the harsher conditions of the upgrade. In the following, the main changes introduced for the upgraded detector are described. Particular focus is given to the tracking system and the RICH detectors.

At higher luminosity, the particle flux increases dramatically in the regions close to the beam axis, therefore a major upgrade is foreseen for the whole LHCb tracking system.

\begin{figure}[!ht]
\centering
 \subfigure[New VELO silicon pixel sensor layout]{\includegraphics[width=6.3cm]{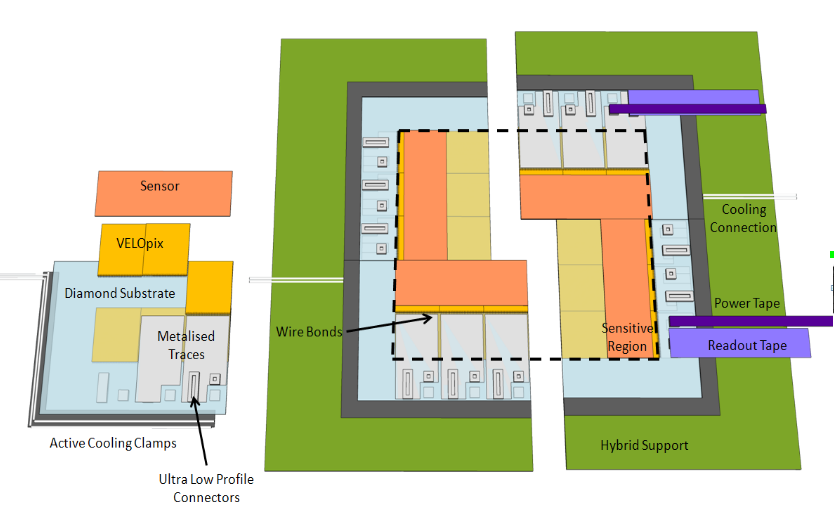}\label{fig:VELOa}}
{\hspace{1cm}}
\subfigure[Micro-channel cooling technology]{\includegraphics[width=6.3cm]{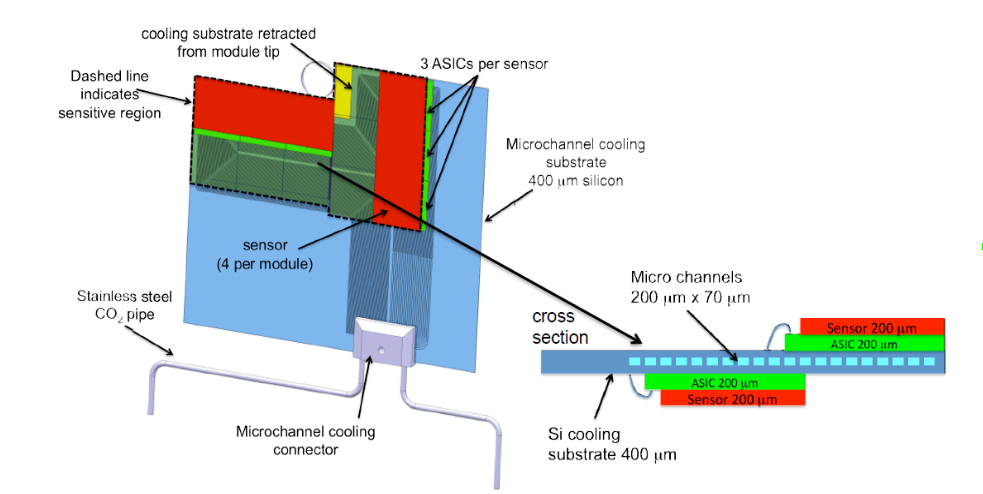}\label{fig:VELOb}}
\caption{The VELO sensors in the upgraded LHCb.}
\label{fig:VELO}
\end{figure}

The current VELO is based on semicircular silicon-strip sensors arranged in two rows that close around the interaction regions during data taking.
While the moving layout will be kept, the baseline choice for the upgrade consists of silicon pixel sensors with an aggressive micro-channel cooling system. The new VELO sensor layout and the micro-channel cooling scheme are shown in \figurename{~\ref{fig:VELO}}. The sensor choice is driven the necessity to reduce the occupancy allowing for a faster track reconstruction and low fake-track rate.

The current Trigger Tracker (TT) will be replaced by the Upstream Tracker (UT). The UT is currently being designed to have a lower material budget (less than $5\%\,X_0$), and to have higher granularity and extended angular coverage compared with the TT.

\begin{figure}[!ht]
\centering
 \includegraphics[width=10cm]{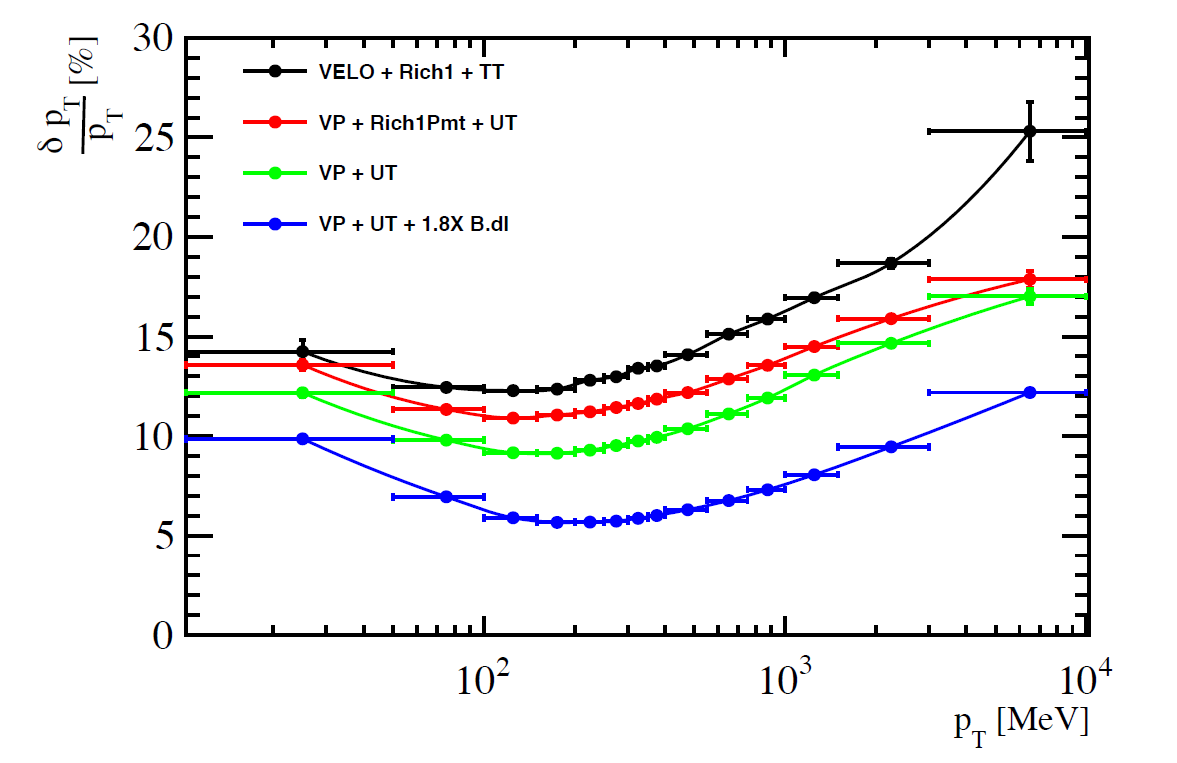}
\caption{Transverse momentum resolution for tracks reconstructed using only information from the vertex detector and the upstream tracker. The performance of the current LHCb detector is shown in black, and the baseline upgrade configuration in green.}
\label{fig:VELUT}
\end{figure}

A comparison of the performance for tracks reconstructed using only information from the VELO and the UT (TT), the so-called \textit{upstream tracks}, in the current detector and in the upgrade scenario, is shown in \figurename{~\ref{fig:VELUT}}.
It is noted that combination of information from the upgraded VELO and UT tracking leads to a considerable improvement in $p_T$ resolution compared with the current VELO+TT.

\begin{figure}[!ht]
\centering
 \includegraphics[width=15cm]{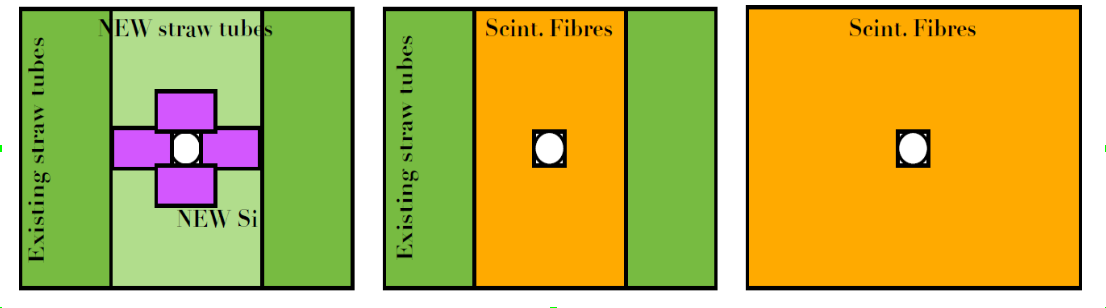}
\caption{Options for the replacement of the current downstream tracking stations. From left to right: replacement of the silicon-strip detector and straw tubes in the central region (outer straw tubes are kept), scintillating fibres detector only in the central region (outer straw tubes are kept), entire downstream tracking station using scintillating fibres technology (baseline).}
\label{fig:SciFi}
\end{figure}

The high track multiplicity in the central region also drives the upgrade of the current downstream tracking stations, located between the dipole magnet and the RICH2 detector. Several detector technologies are currently under study, with the baseline choice being the replacement of the entire inner tracking system (composed of a silicon strip tracker in the inner region and a straw tube outer tracker) with a design known as the \textit{Sci-Fi} detector (see \figurename{~\ref{fig:SciFi}}). The Sci-Fi detector exploits scintillating fibres as the active material. The scintillation light from the fibres is read-out by silicon-based photo-multipliers.

The current RICH system is composed of two detectors, RICH1 and RICH2, located upstream and downstream of the dipole magnet, respectively.
In order to cover a wide momentum range, three radiators are used: aerogel (solid) and $\mathrm{C_4F_{10}}$ (gaseous) in RICH1, and $\mathrm{CF}_4$ in RICH2.
In the upgrade, due to increased occupancy the aerogel, which covers the low momentum range $1\textendash10\,\mathrm{GeV}/c$, will be removed.
Moreover, the current Hybrid-PhotoDetectors will be replaced by  Multi-Anode photo-multipliers which will require new front-end electronics. The optics of both RICH1 and RICH2 will also be optimised.

\section{Prospects for charm physics}
LHCb has a broad upgrade physics programme of which charm measurements are an important part. The large charm production cross-section at $\sqrt{s}=7\,\mathrm{TeV}$, recently measured at LHCb \cite{Aaij:2013mga}, is predicted to increase by a factor of 1.8 at $\sqrt{s}=14\,\mathrm{TeV}$. Exploratory studies indicate that improvements in the trigger strategy could provide an increase of a factor two for the trigger efficiency on charm hadronic decays. The improvement is even more pronounced in multibody decays. In the upgrade era, the charm signal yield is expected to increase by a factor of about 3.6 per $\mathrm{fb}^{-1}$. Since the integrated luminosity recorded per year is expected to also  increase by a factor 3.5 per year, the total charm yield per year could increase by one order of magnitude.

\subsection{Production and spectroscopy}
Charm production and spectroscopy are very active areas of research in LHCb. Recent studies of double-charm production observed as double-charmonium, charmonium and open charm, and double open charm \cite{Aaij:2012dz} can in principle be extended to simultaneous charmonium and bottomonium production in the upgrade era.
The search for new $D_{sJ}$ states \cite{Aaij:2013sza} will also benefit enormously from an increased statistics.
Improvements are also expected in studies of $\chi_{c(1,2,3)}$ production,  $J/\psi$ polarisation, and charmed and doubly charmed baryons.

\subsection{Rare decays}
\begin{figure}[!ht]
\centering
 \subfigure{\includegraphics[width=12cm]{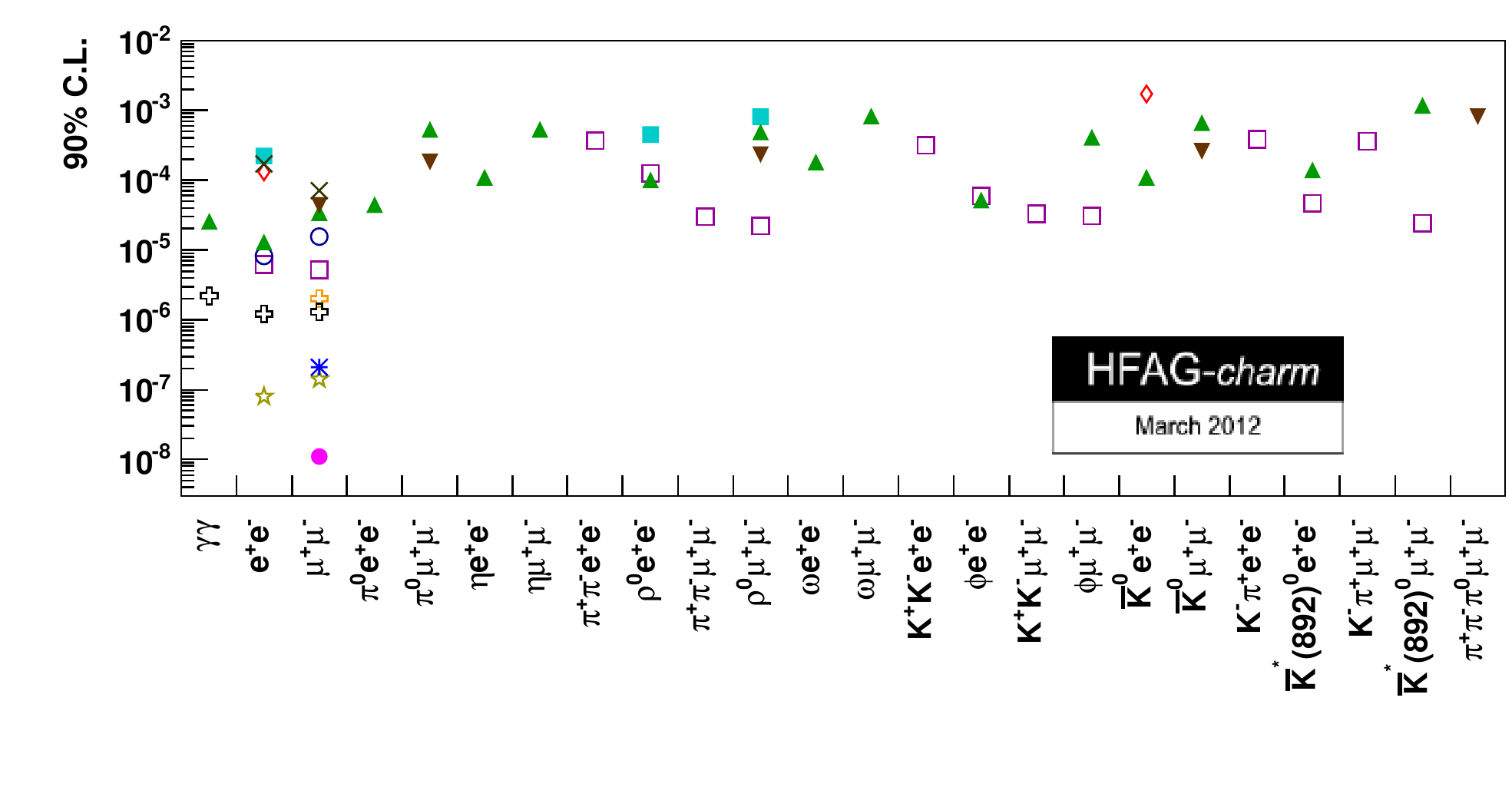}
 \includegraphics[width=3cm]{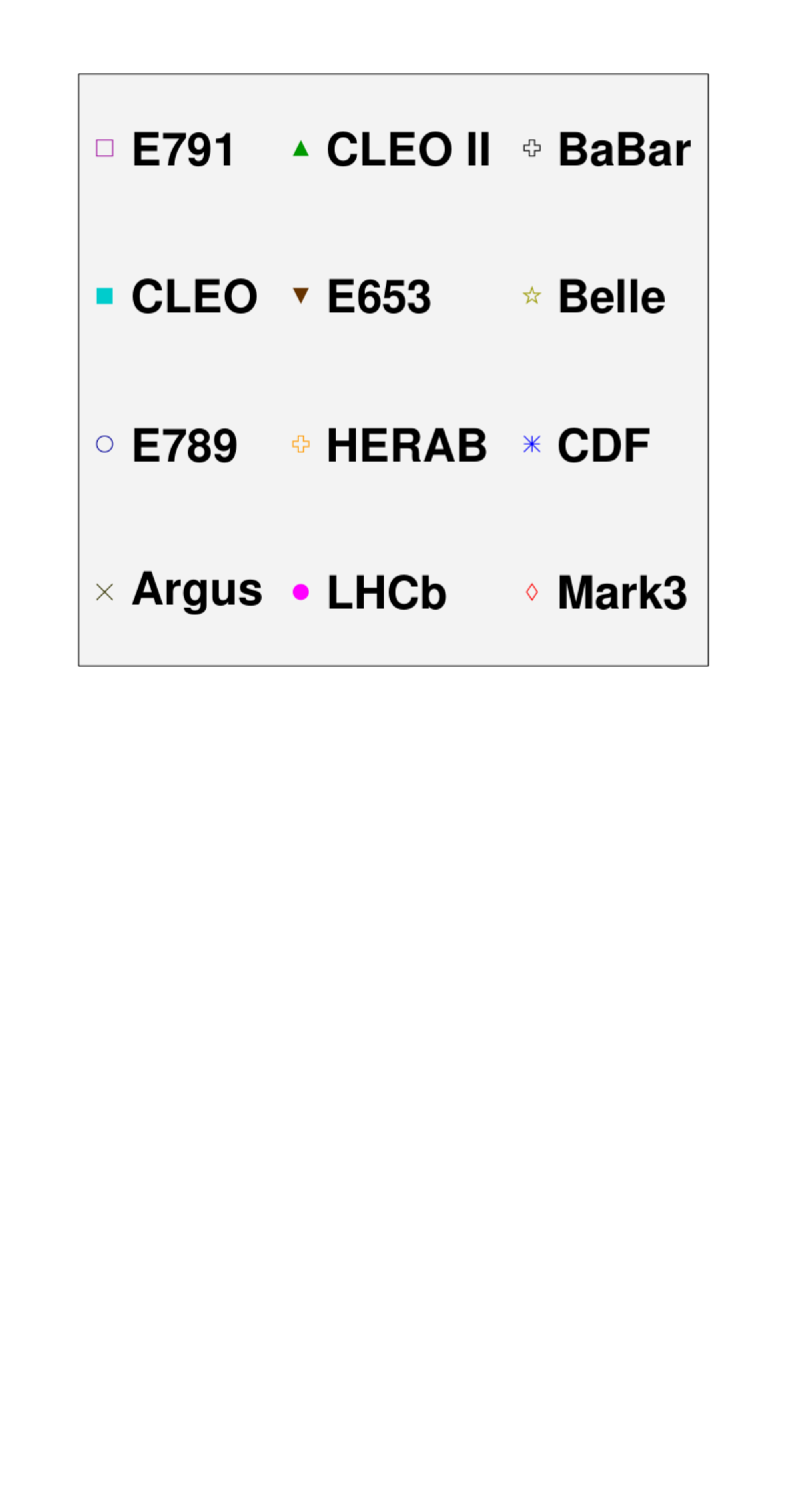}}
\caption{Current limits on rare $D^0$ decays \cite{Amhis:2012bh}.}
\label{fig:hfagrare}
\end{figure}

Charm rare decays are very powerful means to search for new mediators and couplings. The current overview of for $D^0$ decays is shown in \figurename{~\ref{fig:hfagrare}}. LHCb results on $D^0\to\mu^+\mu^-$ \cite{Aaij:2013cza} (see \figurename{~\ref{fig:hfagrare}}) and multibody decays, such as $D^+_{(s)}\to \pi^+ \mu^+\mu^-$ and $D^+_{(s)}\to \pi^- \mu^+\mu^+$ \cite{Aaij:2013sua} and $D^0\to \pi^+\pi^- \mu^+\mu^-$ \cite{Aaij:2013uoa}, are already available and improved previous measurements by one or two orders of magnitude.
Multibody decays may proceed via an intermediate resonance, e.g. $D^+_{(s)}\to \pi^+ \phi$ and then $\phi\to\mu^+\mu^-$. In this context the rare decay searches mentioned above are for the non-resonant modes. However, the resonant modes are themselves of interest for an angular analyses. There is particular interest in the study of forward-backward asymmetries, T-odd correlations and near-resonance effects. Decay modes with intermediate resonances in the dimuon mass can already be seen in the current LHCb data sample. The statistical precision required for angular analyses is expected to be available at the end of the LHCb upgrade. 

It is noted that hadronic modes are a dangerous background to rare decay searches, having a branching fractions $\mathcal{O}(10^{6})$ larger than typical predictions for electroweak $D$ meson decays in the SM. While this background is greatly reduced with information from the muon chambers, decays in flight of high momentum pions into muons can easily mimic a genuine muon directly from a $D$ decay in such a way that hadronic decays become an irreducible background. Since the discriminating power is currently reaching a limit, improvements in the muon identification in the upgrade are one key ingredient for the progress in this area.

\subsection{Mixing}
Charm mixing is already established by a series of complementary measurements although considerable improvements are still needed in the precision with which the mixing parameters $x$ and $y$ are known. The LHCb collaboration, analysing data collected during the 2011 run only, made the first single measurement to exclude the no-mixing hypothesis to a level above five standard deviations \cite{Aaij:2012nva}. The analysis, based on the study of the time-dependent ratio between wrong- (WS) and right-sign (RS) $D^0\to K^\mp\pi^\pm$, is a perfect demonstration of the LHCb's statistical power. The updated analysis based on the complete Run 1 LHCb data sample ($3\,\mathrm{fb}^{-1}$), also contains the most precise determination of the mixing parameters $x'$ and $y'$ and a search for $CP$ violation \cite{Aaij:2013wda}.

Another observable which give access to the mixing parameters is $y_{CP}$, defined as the ratio between the effective lifetime for decays to $CP$-even eigenstate ($K^+K^-$ or $\pi^+\pi^-$) and Cabibbo-favoured decays to the $CP$-mixed final state $K^-\pi^+$. The current measurement from LHCb, based on a small data sample collected in 2010, proves the feasibility of the measurement at hadron machines \cite{Aaij:2011ad}. An updated measurement, which uses the 2011 dataset, is in progress. The large yields available in the upgrade will allow a more refined treatment of backgrounds that will reduce the systematic uncertainty affecting the measurement.

Other mixing measurement under study within the LHCb collaboration include:
\begin{itemize}
 \item $x^2+y^2$ using the time integrated WS/RS ratio of $D^0\to K^+\mu^- \nu$ decays
 \item Direct access to $x$ and $y$ via a time-dependent Dalitz plot measurement of $D^0\to K_Shh$ decays
\item Access to $x''^2$ and $y''$ via a time-dependent WS/RS Dalitz plot measurement of $D^0\to K^+\pi^+\pi^0$
\end{itemize}

The sensitivities expected for several mixing observables, extrapolated to an integrated luminosity of $50\,\mathrm{fb^{-1}}$ (note that the expected luminosity has increased since these estimates were made in Ref.{\cite{Bediaga:2012py}}), are summarised in \tablename{~\ref{tab:predictions1}}.
\begin{table}[!ht]
\begin{center}
\begin{tabular}{l|cc}
 Decay & Observable & Exp sensitivity $[\times10^{-3}]$ (stat only)\\
\hline
$D^0\to KK$ & $y_{CP}$ & 0.04\\
$D^0\to \pi\pi$ & $y_{CP}$ & 0.08\\
$D^0\to K^+\pi^-$ & $x'^2$,$y'$ & 0.04,0.1\\
$D^0\to K_S\pi\pi$ & $x$,$y$ & 0.15,0.1\\
$D^0\to K^+\mu^-\nu$ & $R_M=x^2+y^2$ & 0.0001\\
\hline
\end{tabular}
\caption{Projection of statistical sensitivities for mixing observables with $50\,\mathrm{fb^{-1}}$ \cite{Bediaga:2012py}.}
\label{tab:predictions1}
\end{center}
\end{table}

\subsection{Indirect $CP$ violation}
As well as the $CPV$ search in the time-dependent wrong-sign $D^0\to K^+\pi^-$ decay mentioned previously, LHCb is carring out a search for indirect $CP$ violation in the charm sector through the measurement of $A_\Gamma$ \cite{Aaij:2013ria}. The parameter $A_\Gamma$, defined as the asymmetry between the effective lifetimes of $D^0$ decays into a $CP$ eigenstate, is an almost clean measurement of indirect $CP$ violation and can expressed as
\begin{equation}
A_\Gamma = \frac{1}{2}(A_m+A_d)y\cos\phi-x\sin\phi \approx -a^{ind}_{CP}  -a^{dir}_{CP}y_{CP},
\label{eq:indcpv}
\end{equation}
where $A_m=1-|q/p|$, $A_d = 1-|A_f/\bar{A}_f|$ and $\phi$ is the relative $CP$ violating phase between $q/p$ and $\bar{A}_f/A_f$.
In Eq.~\ref{eq:indcpv} it is manifest that this measurements benefits from a precise determination of the mixing parameters $x$ and $y$, which are expected to be constrained at a $10^{-4}$ level in the upgrade. Since the overall precision on $A_\Gamma$ at the end of the upgraded LHCb data-taking is expected to be better than $10^{-4}$, a precision independent measurement of the direct $CP$ violating component is necessary to probe the SM prediction for $A_\Gamma$ which is set to about $10^{-4}$. 

In addition to the mixing parameters, $D^0\to K_Sh^+h^-$ decays which give also access to $CP$ violating quantities such as $|q/p|$ and $\phi$, making this a ``golden-channel'' for the LHCb upgrade. These parameters are accessible via a the time-dependent evolution in the $K_S\pi\pi$ Dalitz plane. Two strategies are possible: an unbinned, model-dependent measurement in which a full amplitude fit is performed, and a model-independent measurement that instead uses prior experimental measurements of the average strong phase difference in regions of the Dalitz-plot (e.g. from CLEOc and BESIII).
Although such decays suffer from a relatively low reconstruction efficiency in LHCb, mainly due to the $K_S$ long lifetime, precise measurements of $x$, $y$, $q/p$ and $\phi$ can already be performed with the existing data samples and will be greatly improved in the upgrade.

\subsection{Direct $CP$ violation}
Measurements of direct $CP$ violation are challenge for experiments at hadron colliders. In fact, several sources of asymmetry can bias the measurement such as the production asymmetry present in proton-proton collisions. Moreover, analyses can be affected by detection asymmetry biases.
Therefore, independent measurements of production and detection asymmetries are a crucial ingredient for direct $CP$ violation searches in charm. These measurements are currently being performed within the LHCb collaboration \cite{Aaij:2012cy,LHCb:2012fb,LHCb-CONF-2013-023} and will be pursued in the upgrade phase.

It is interesting to note that if detection and production asymmetries are small, observables can be constructed in which they cancel at the first order. This fact is exploited in the measurement of $\Delta A_{CP}=A_{CP}(K^+K^-)-A_{CP}(\pi^+\pi^-)$ in prompt \cite{LHCb-CONF-2013-003} and semileptonic \cite{Aaij:2013bra} decays performed by LHCb. The improved detector and the larger statistics of the LHCb upgrade are therefore vital to reduce the statistical and systematic uncertainties and shed light on the still unclear picture of direct $CP$ violation in the charm sector.

The sensitivities for several direct $CP$ violating observables are given in \tablename{~\ref{tab:predictions2}}, assuming an integrated luminosity of $50\,\mathrm{fb^{-1}}$.
\begin{table}[!ht]
\begin{center}
\begin{tabular}{l|cc}
 Decay & Observable & Exp sensitivity $[\times10^{-3}]$ (stat only)\\
\hline
$D^0\to KK,\pi\pi$ & $\Delta A_{CP}$ & 0.15\\
$D^+\to K_S K^+$ & $A_{CP}$ & 0.1\\
$D^+\to K^-K^+\pi^+$ & $A_{CP}$ & 0.05\\
$D^+\to \pi\pi\pi$ & $x$,$y$ & 0.08\\
$D^+\to hh\pi$ & $CPV$ in phases & $(0.01-0.10)^\circ$\\
$D^+\to hh\pi$ & $CPV$ in fractions & $0.1-1.0$\\
\hline
\end{tabular}
\caption{Projection of statistical sensitivities for $CP$ observables with $50\,\mathrm{fb^{-1}}$ \cite{Bediaga:2012py}.}
\label{tab:predictions2}
\end{center}
\end{table}

\section{Conclusions}
The LHCb detector is performing excellently and is already exceeding its design expectations confirming the feasibility of charm physics at hadron colliders. The collaboration is active in many complementary analysis in the charm sector, and in particular sub-percent measurements of several $CP$ quantities are expected to be already available before the upgrade and will reach or even exceed the current theoretical precision after the upgrade. In the upgrade era, these studies will be further improved thanks to the increased statistics and the improvements in the hardware and software infrastructure.
In addition, the upgraded LHCb detector has tremendous potential for new measurements in charm rare decays, production and spectroscopy.
In parallel, ongoing efforts are focused on reducing possible sources of systematic uncertainties that may limit the LHCb scope.
Further, detailed and information on the LHCb upgrade is reported in \cite{CERN-LHCC-2011-001,Bediaga:1443882}.

\section*{Acknowledgements}
 
The text below are the acknowledgements as approved by the collaboration
board. Extending the acknowledgements to include individuals from outside the
collaboration who have contributed to the analysis should be approved by the
EB and, if possible, be included in the draft of first circulation.
 
\noindent We express our gratitude to our colleagues in the CERN
accelerator departments for the excellent performance of the LHC. We
thank the technical and administrative staff at the LHCb
institutes. We acknowledge support from CERN and from the national
agencies: CAPES, CNPq, FAPERJ and FINEP (Brazil); NSFC (China);
CNRS/IN2P3 and Region Auvergne (France); BMBF, DFG, HGF and MPG
(Germany); SFI (Ireland); INFN (Italy); FOM and NWO (The Netherlands);
SCSR (Poland); MEN/IFA (Romania); MinES, Rosatom, RFBR and NRC
``Kurchatov Institute'' (Russia); MinECo, XuntaGal and GENCAT (Spain);
SNSF and SER (Switzerland); NAS Ukraine (Ukraine); STFC (United
Kingdom); NSF (USA). We also acknowledge the support received from the
ERC under FP7. The Tier1 computing centres are supported by IN2P3
(France), KIT and BMBF (Germany), INFN (Italy), NWO and SURF (The
Netherlands), PIC (Spain), GridPP (United Kingdom). We are thankful
for the computing resources put at our disposal by
Yandex LLC (Russia), as well as to the communities behind the multiple open
source software packages that we depend on.


\begin{thebibliography}{99}


\bibitem{Aaij:2013mga}
  R.~Aaij {\it et al.}  [LHCb Collaboration],
  Nucl.\ Phys.\ B {\bf 871} (2013) 1

\bibitem{Alves:2008zz}
LHCb collaboration, A.~A. Alves~Jr. {\em et~al.},
  {JINST {\bf 3} (2008) S08005}

\bibitem{LHCb-DP-2012-003}
M.~Adinolfi {\em et~al.}, 
{Eur.\ Phys.\ J.\  {\bf C73} (2013) 2431}, 

\bibitem{LHCb-DP-2012-002}
A.~A. Alves~Jr. {\em et~al.}, 
  {JINST {\bf 8} (2013) P02022}

\bibitem{Aaij:2012dz}
  R.~Aaij {\it et al.}  [LHCb Collaboration],
  JHEP {\bf 1206} (2012) 141

\bibitem{Aaij:2013sza}
  R.~Aaij {\it et al.}  [LHCb Collaboration],
  JHEP {\bf 1309} (2013) 145

\bibitem{Aaij:2013cza}
  R.~Aaij {\it et al.}  [LHCb Collaboration],
  Phys.\ Lett.\ B {\bf 725} (2013) 15

\bibitem{Aaij:2013sua}
  R.~Aaij {\it et al.}  [LHCb Collaboration],
  Phys.\ Lett.\ B {\bf 724} (2013) 203

\bibitem{Aaij:2013uoa}
  R.~Aaij {\it et al.}  [LHCb Collaboration],
  arXiv:1310.2535 [hep-ex].

\bibitem{Amhis:2012bh}
  Y.~Amhis {\it et al.}  [Heavy Flavor Averaging Group Collaboration],
  arXiv:1207.1158 [hep-ex].

\bibitem{Aaij:2012nva}
  R.~Aaij {\it et al.}  [LHCb Collaboration],
  Phys.\ Rev.\ Lett.\  {\bf 110} (2013) 10,  101802

\bibitem{Aaij:2013wda}
      R.~Aaij {\it et al.}  (LHCb Collaboration),
      arXiv:1309.6534 [hep-ex].


\bibitem{Aaij:2011ad}
  R.~Aaij {\it et al.}  [LHCb Collaboration],
  JHEP {\bf 1204} (2012) 129

\bibitem{Bediaga:2012py}
  R.~Aaij {\it et al.}  [LHCb Collaboration],
  Eur.\ Phys.\ J.\ C {\bf 73} (2013) 2373


\bibitem{Aaij:2013ria}
  R.~Aaij {\it et al.}  [LHCb Collaboration],
  arXiv:1310.7201 [hep-ex].


\bibitem{Aaij:2012cy}
  RAaij {\it et al.}  [LHCb Collaboration],
  Phys.\ Lett.\ B {\bf 713} (2012) 186

\bibitem{LHCb:2012fb}
  RAaij {\it et al.}  [LHCb Collaboration],
  Phys.\ Lett.\ B {\bf 718} (2013) 902
  [arXiv:1210.4112 [Unknown]].
\bibitem{LHCb-CONF-2013-023}
  R.~Aaij {\it et al.}  [LHCb Collaboration],
   LHCb-CONF-2013-023


\bibitem{LHCb-CONF-2013-003}
   R.~Aaij {\it et al.}  [LHCb Collaboration],
   LHCb-CONF-2013-003


\bibitem{Aaij:2013bra}
  R.~Aaij {\it et al.}  [LHCb Collaboration],
  Phys.\ Lett.\ B {\bf 723} (2013) 33



\bibitem{CERN-LHCC-2011-001}
  CERN, Technical report,
  CERN-LHCC-2011-001. LHCC-I-018

\bibitem{Bediaga:1443882}
  I.~Bediaga {\it et al.} [LHCb Collaboration],
  Technical report,
  CERN-LHCC-2012-007. LHCb-TDR-12

    



\end{thebibliography}
\end{document}